\documentclass[aps,pra,twocolumn,groupedaddress,showpacs]{revtex4}
\usepackage{graphicx}
\usepackage[ansinew]{inputenc}
\usepackage{array}
\usepackage{color}
\usepackage{amsmath}
\usepackage{amsxtra}
\usepackage{amssymb}
\usepackage{latexsym}
\usepackage{dsfont}
\begin{document}
\title{Detecting maximally entangled states without making 
the Schmidt decomposition}
\author{M. Bhattacharya}
\affiliation{B2 Institute, Department of Physics and College of
Optical Sciences, The University of Arizona, Tucson, Arizona
85721}

\date{\today}

\begin{abstract}
The bipartite entanglement of a pure quantum state is 
known to be characterized by its Schmidt decomposition. 
In particular the state is maximally entangled when 
all the Schmidt coefficients are equal. We point out
a convenient method which always yields a single 
analytical condition for the state to be maximally
entangled, in terms of its expansion coefficients 
in any basis. The method works even when the Schmidt 
coefficients cannot be calculated analytically, and
does not require their calculation. As an example this
technique is used to derive the Bell basis for a system of
two qubits. In a second example the technique shows a 
particular state to \textit{never} be maximally entangled, 
a general conclusion that cannot be reached using the Schmidt 
decomposition. 
\end{abstract}


\maketitle

\section{Introduction}
\label{sec:intro} 
Entanglement or the existence of quantum correlations
between physical systems is currently of great interest 
both theoretically and experimentally, see Ref.~\cite{Nielsenbook}
and references therein. From the theoretical point of view 
the study of entanglement is shedding light on the basic 
nature of quantum mechanics \cite{horodeckirmp}, and 
providing surprising connections to other areas, such 
as the study of black holes \cite{kallosh2006}. From a 
practical point of view it has been realized that 
entanglement is a resource that can enable information 
processing tasks that turn out to be impossible or 
inefficient when tackled by classical machines 
\cite{ExpQIbook}. These capabilities include teleportation, 
cryptography, and computation \cite{Nielsenbook}. 

In this article we will consider the entanglement of pure
bipartite states, i.e. states which describe globally pure
quantum mechanical systems which have been partitioned into 
two subsystems. As is known, the Schmidt decomposition of a 
pure bipartite state provides a qualitative measure of 
its entanglement \cite{horodeckirmp,Nielsenbook}. On the 
other hand the von Neumann entropy of either subsystem 
provides a unique quantitative measure of the degree of 
entanglement of the whole quantum state \cite{horodeckirmp,
Nielsenbook}. 

In particular we will consider \textit{maximally}
entangled states i.e. those states of a quantum system 
which contain the largest amount of entanglement possible. 
Since entanglement is a resource for information processing,
these states are of particular importance. Indeed protocols 
for quantum key distribution, dense coding and teleportation 
rely on maximally entangled states \cite{horodeckirmp}. 
In view of the inevitable noise and decoherence that 
accompanies an experiment, maximally entangled states 
also provide a benchmark for other, less entangled, states, 
as in the process of distillation \cite{Nielsenbook}. 
The Bell states are perhaps the best known example of 
maximally entangled bipartite pure states. 

How do we know if a pure bipartite state is maximally 
entangled ? We can decide using either of the measures 
mentioned above. If the state is entangled maximally 
all the coefficients in the Schmidt decomposition are 
equal; also the reduced von Neumann entropy is maximised 
to its upper bound log $d$, where $d$ is the dimension 
of the smaller partition. However if $d>4$ neither of 
these measures can provide analytical conditions (on the
expansion coefficients of the state in some basis), and 
have to be computed numerically.

In this article we point out an existing technique from 
algebra that can provide a single analytical condition 
(for arbitrary but finite $d$) that the coefficients 
(in any basis) of a maximally entangled state have to 
obey. The method therefore is capable of deciding quite
generally if a certain state is maximally entangled or 
not. We demonstrate below how this method shows a certain
state can never be maximally entangled, a general conclusion
that cannot be reached using the Schmidt decomposition or
the von Neumann entropy. In addition the method can also 
be used to obtain the maximally entangled basis given the 
dimensions of the two subsystems. As an example below we 
use the method to derive the Bell basis for a system of 
two qubits. 

The rest of the article is arranged as follows. In 
Section ~\ref{sec:gform} we outline the general algorithm. 
In Section ~\ref{sec:bell} we use the technique to derive 
the Bell basis for a system of two qubits providing also a
detailed commentary on our method, and its simple 
implementation in \textit{Mathematica}. In Section ~\ref{sec:five}
we consider a case which cannot be solved analytically using
the Schmidt decomposition or the von Neumannn entropy. 
Section ~\ref{sec:conc} provides a Conclusion.

\section{General formalism}
\label{sec:gform}
The steps of our technique are as follows:
\begin{enumerate}
 \item Begin with a quantum state $|\psi \rangle$.
 \item Find the density matrix $\rho=|\psi\rangle 
       \langle \psi |.$
 \item Trace over the states of one subsystem, 
       say $A$, to find the reduced density matrix 
       $\rho_B=\mathrm{Tr}_A \rho$.
 \item Find the characteristic polynomial $P[\rho_B,x]$
       of this matrix with respect to a dummy variable
       $x$.
 \item Find the subdiscriminant sequence $D_q [P]$, 
       of the polynomial $P$, where $q=1,2\ldots d$ \cite{Basubook}.
 \item If the last but one member of the sequence, i.e. $D_{d-1} [P]$, 
       equals zero then $|\psi \rangle$ is maximally
       entangled.
\end{enumerate}

The basic tool we use is the subdiscriminant sequence 
of the characteristic polynomial of the reduced density 
matrix of either subsystem. The subdiscriminant sequence 
of any polynomial can be found in textbooks on algebraic 
geometry such as Ref.~\cite{Basubook}. A detailed 
introduction for physicists has been provided in 
Ref.~\cite{mbcr122007} and will not be repeated here, 
although we will discuss a few important points. In general 
the sequence contains $d$ members. The various members 
denote the number of repeating zeros of the polynomial. 
For example the first member of the sequence is the 
discriminant of the polynomial $P$
\begin{eqnarray}
\begin{array}{l}
\label{eq:Disc} D_{1}[P]=
\displaystyle\prod_{i<j}^{d}(\lambda_{i}-\lambda_{j})^{2},\\
\end{array}
\end{eqnarray}
where $\lambda_i$ are the roots of the polynomial $P$. Since
density matrices are Hermitian, the $\lambda_i$ are all real.
From Eq.~(\ref{eq:Disc}) we can see that $D_1[P]$ vanishes 
whenever \textit{two or more} roots of $P$ are equal. 

For the purpose of this article the most important member of 
the subdiscriminant sequence is its second to last entry
\begin{equation}
 \label{eq:Dlastbutone}
D_{d-1}[P]= \displaystyle\sum_{i<j}^{d}(\lambda_{i}-\lambda_{j})^{2},
\end{equation}
which equals zero only if \textit{all} the eigenvalues of the 
reduced density matrix are equal. In turn this implies the 
equality of the Schmidt coefficients and maximal entanglement 
of the state. It is an important fact that $D_{1}[P],\, D_{d-1}[P]$, 
and actually the whole subdiscriminant sequence $D_{q}[P]$ can 
always be found analytically in terms of the coefficients of 
the polynomial $P$ \cite{mbcr122007}. Significantly this does 
not require calculation of the eigenvalues $\lambda_i$, i.e. 
one need not make the Schmidt decomposition \cite{Basubook}. 

Generally $D_{d-1}[P]$, which has to vanish for a maximally 
entangled state, is itself a polynomial in the coefficients 
of expansion of the given quantum state written in an arbitrary 
basis. It follows that from a broader perspective the 
present technique maps the problem of finding maximally 
entangled states of a bipartite system to that of finding the 
roots of a multivariate polynomial. In order to expose the 
working details of our procedure we present two examples. 
We first rederive the familiar Bell states using our method; 
then we consider a more involved example which cannot be 
solved analytically using Schmidt decomposition or the von 
Neumann entropy.
\section{A $2 \times 2$ system : the Bell basis}
\label{sec:bell}
We consider a system which is divided into two subsystems
$A$ and $B$. Each subsystem contains a qubit which can 
exist in a superposition of the states $|0\rangle$ and 
$|1\rangle$. An arbitrary unnormalized pure quantum 
state of this system can be written as
\begin{equation}
 \label{eq:qubits}
|\psi \rangle = p |00\rangle + q |11\rangle + r |10\rangle + s |01\rangle.
\end{equation}
Here we have used the product basis, which is often a
convenient one to use, although for the method to be 
demonstrated $|\psi \rangle$ can be expressed in any basis.
For the particular task of deriving the Bell states 
we have chosen the coefficients $(p,q,r,s)$ to be 
all real.

The density matrix corresponding to the state in 
Eq.~(\ref{eq:qubits}) can be written easily and its 
trace over the first qubit yields the reduced density
matrix
\begin{equation}
 \label{eq:reddens}
\rho_B = \left(
\begin{array}{ll}
 p^2+r^2& ps+qr \\
 ps+qr & q^2 + s^2 \\
\end{array}
\right),
\end{equation}
where we have used the representation
\begin{equation}
 \label{eq:qubrep}
|0 \rangle \langle 0| = 
\left(
\begin{array}{ll}
 1 & 0 \\
 0 & 0 \\
\end{array}
\right),
|1 \rangle \langle 1| = 
\left(
\begin{array}{ll}
 0 & 0 \\
 0 & 1 \\
\end{array}
\right),
|0 \rangle \langle 1| = 
\left(
\begin{array}{ll}
 0 & 1 \\
 0 & 0 \\
\end{array}
\right),
\end{equation}
etc. The characteristic polynomial of $\rho_B$ 
[Eq.~(\ref{eq:reddens})] is
\begin{equation}
 \label{eq:cpqubits}
P[\rho_B,x]=x^2-\left(p^2+q^2+r^2+s^2\right) x + (p q-r s)^2,
\end{equation}
where $x$ is a dummy variable. Clearly $P$ is 
quadratic in $x$ and has two roots which are 
non-zero in general. They are the eigenvalues of 
$\rho_B$ and yield the Schmidt coefficients 
\cite{horodeckirmp}; as promised, we will not 
calculate them.

The \texttt{Subresultants}$[P, P', x]$ 
function in \textit{Mathematica} directly yields
the subdiscriminant sequence with the polynomial $P$ and its 
derivative $P'$ with respect to $x$ as inputs. We note that 
\textit{Mathematica} requires that the coefficient of the 
highest power of $x$ in $P$ as well as  $P'$ to be 1. This 
can be arranged easily. Also in the general case the number 
of terms in the sequence equals the degree of $P$, i.e. $d$. 
In the present example therefore there are two terms in the 
sequence. The penultimate term in the subdiscriminant sequence 
of $P$ in Eq.~(\ref{eq:cpqubits}) is the discriminant of $P$, 
$D_{1}[P]$, which equals zero whenever the two roots of $\rho_B$ 
coincide \cite{mbcr122007,Basubook}. We find
\begin{equation}
 \label{eq:discqubits}
D_{1}[P]=\left[(p+q)^2+(r-s)^2\right] \left[(p-q)^2+(r+s)^2\right].
\end{equation}
If $| \psi \rangle$ is to be maximally entangled $D_{1}[P]=0.$ 
We can extract the Bell basis using this criterion. 
Specifically, we can see that $(p=q=0, r=\pm s)$ is 
a solution set and yields, using Eq.~(\ref{eq:qubits}), 
the states
\begin{equation}
 \label{eq:Bell12}
|\psi_{1,2} \rangle = \frac{|10 \rangle \pm |01 \rangle}{\sqrt{2}},
\end{equation}
after normalization. Similarly, $(p=\pm q, r=s=0,)$ is a solution 
set and yields the states
\begin{equation}
 \label{eq:Bell34}
|\psi_{3,4} \rangle = \frac{|00 \rangle \pm |11 \rangle}{\sqrt{2}},
\end{equation}
after normalization. As is well known, the states 
$|\psi_{1,2,3,4}\rangle$ constitute the Bell basis 
\cite{Nielsenbook}. Any state, and therefore any 
maximally entangled state can be expressed in this 
basis. For example, the solution set $(p=-q, r=s)$ 
yields the (unnormalized) maximally entangled state
\begin{equation}
 \label{eq:qubitothers}
|\psi \rangle= p\sqrt{2}|\psi_4 \rangle + r\sqrt{2}|\psi_1 \rangle.
\end{equation}
We note that $(p,q,r,s)$ can be complex in general; 
however the analysis in that case is not very different
from that presented here (also see below).

\section{A $5 \times 5$ system}
\label{sec:five}
The superiority of the method proposed in this article
over that of Schmidt decomposition or von Neumann entropy 
calculation becomes clearer for systems of higher dimension, 
i.e. qudits. Physically qudits are of interest as some of 
the proposed candidates for quantum computation possess $d>2$, 
such as a molecule with ro-vibrational states \cite{shapiro2003} 
and an alkali atom with hyperfine-Zeeman levels \cite{jessen2001}. 

Here we consider a system with two parts each containing 
a 5-level system with states $|0 \rangle,
|1 \rangle, |2 \rangle, |3 \rangle$, and $|4 \rangle$.
Specifically, we consider the unnormalized state 
\begin{eqnarray}
 \label{eq:five}
| \psi \rangle \,\,=&& |0 2 \rangle +2|1 0 \rangle +|2 0 \rangle 
                +|2 1 \rangle +| 2 2 \rangle + | 2 3 \rangle  \nonumber \\ 
                &&+|2 4 \rangle + 3 | 3 3 \rangle + p | 4 4 \rangle,
\end{eqnarray}
where $p$ is the only unknown \textit{real} coefficient 
of expansion in the product basis. Tracing the density 
matrix over the five levels of system $A$, we obtain
\begin{equation}
 \label{eq:rhobfive}
\rho_B=\left(
\begin{array}{ccccc}
 5 & 1 & 1 & 1 & 1 \\
 1 & 1 & 1 & 1 & 1 \\
 1 & 1 & 2 & 1 & 1 \\
 1 & 1 & 1 & 10 & 1 \\
 1 & 1 & 1 & 1 & p^2+1
\end{array}
\right),
\end{equation}
where we have used a matrix representation analogous to
Eq.~(\ref{eq:qubrep}). The characteristic polynomial
of $\rho_B$ [Eq.~(\ref{eq:rhobfive})] is 
\begin{eqnarray}
 \label{eq:cpqudits}
P[\rho_B,x]=-x^5+\left(p^2+19\right) x^4-\left(18 p^2+105\right) x^3+ \nonumber \\
\left(91 p^2+183\right) x^2-\left(134 p^2+72\right) x+36 p^2.
\end{eqnarray}
$P$ is a quintic, and not generally solvable analytically 
in terms of radicals \cite{Basubook}. This implies that 
the Schmidt coefficients have to be found numerically. 
However the subdiscriminant sequence can be found 
analytically and its fourth member, apart from an 
irrelevant numerical prefactor, is
\begin{equation}
 \label{eq:subdqudit}
D_{4}[P]=2 p^4-14 p^2+197.
\end{equation}
$|\psi \rangle$ is therefore maximally entangled when $D_{4}[P]=0$,
which yields the four roots
\begin{equation}
 \label{eq:solqubitmax}
p= \pm \left( \frac{7+i\sqrt{345}}{2}\right)^{1/2},\,\,\pm \left(\frac{7-i\sqrt{345}}{2}\right)^{1/2}.
\end{equation}
Since none of these solutions are real, $|\psi \rangle$ 
can \textit{never} be maximally entangled. 

We note that if we had assumed $p$ to be complex in 
Eq.~(\ref{eq:five}), then Eqs.~(\ref{eq:rhobfive}),
~(\ref{eq:cpqudits}), and ~(\ref{eq:subdqudit}) would 
undergo the transform $p^2 \rightarrow |p|^2$.  
The conditions presented in Eq.~(\ref{eq:solqubitmax}) 
would then hold for $|p|$ and would be impossible to achieve
since by definition $|p|$, the modulus of $p$, is real. 
Therefore even if $p$ is allowed to be complex 
$|\psi \rangle$ can never be entangled maximally. We 
also note that the case of Eq.~(\ref{eq:subdqudit}) 
is exceptional in that it is an analytically solvable 
equation, i.e. a quartic, in the variable of interest, 
$p$. This is atypical. Although the subdiscriminant 
sequence can always be obtained analytically, the 
roots of its members typically have to be found numerically. 

The above demonstration, although it uses a somewhat 
arbitrary quantum state, shows the general usefulness 
of the method introduced in this article. The inability 
to maximally entangle the state $|\psi \rangle$ of 
Eq.~(\ref{eq:five}) for \textit{any} value of $p$ cannot 
be established generally by calculating the Schmidt 
coefficients or  the von Neumann entropy.
\section{Conclusion}
\label{sec:conc}
In this article we have pointed out a technique 
that always yields a single analytical condition that
is satisfied by the coefficients of a pure bipartite 
quantum state if it is maximally entangled. The method 
is superior to that of Schmidt decomposition and von 
Neumann entropy for multilevel systems of dimension 
greater than 4. Further, it can also be used to obtain 
the maximally entangled basis for systems of arbitrary
but finite dimension.
\begin{acknowledgments}
This work is supported in part by the US Office of Naval 
Research, by the National Science Foundation and by the 
US Army Research Office. It is a pleasure to thank 
P. Meystre for support and H. Uys, P.-L. Giscard and 
B. McMorran for enlightening discussions. 
\end{acknowledgments}


\begin{thebibliography}{8}
\expandafter\ifx\csname natexlab\endcsname\relax\def\natexlab#1{#1}\fi
\expandafter\ifx\csname bibnamefont\endcsname\relax
  \def\bibnamefont#1{#1}\fi
\expandafter\ifx\csname bibfnamefont\endcsname\relax
  \def\bibfnamefont#1{#1}\fi
\expandafter\ifx\csname citenamefont\endcsname\relax
  \def\citenamefont#1{#1}\fi
\expandafter\ifx\csname url\endcsname\relax
  \def\url#1{\texttt{#1}}\fi
\expandafter\ifx\csname urlprefix\endcsname\relax\def\urlprefix{URL }\fi
\providecommand{\bibinfo}[2]{#2}
\providecommand{\eprint}[2][]{\url{#2}}

\bibitem[{\citenamefont{Nielsen and Chuang}(2000)}]{Nielsenbook}
\bibinfo{author}{\bibfnamefont{M.~A.} \bibnamefont{Nielsen}} \bibnamefont{and}
  \bibinfo{author}{\bibfnamefont{I.~L.} \bibnamefont{Chuang}},
  \emph{\bibinfo{title}{Quantum Computation and Quantum Information}}
  (\bibinfo{publisher}{Cambridge University Press},
  \bibinfo{address}{Cambridge}, \bibinfo{year}{2000}), \bibinfo{edition}{1st}
  ed.

\bibitem[{hor()}]{horodeckirmp}
\bibinfo{note}{R. Horodecki, P. Horodecki, M. Horodecki and K. Horodecki,
  arXiv:quant-ph/0702225v2 (2007), submitted to Reviews of Modern Physics.}

\bibitem[{kal()}]{kallosh2006}
\bibinfo{note}{R. Kallosh and A. Linde, Phys. Rev. D \textbf{73}, 104033
  (2006).}

\bibitem[{\citenamefont{Martini and C.~Monroe}(2002)}]{ExpQIbook}
\bibinfo{author}{\bibfnamefont{F.~D.} \bibnamefont{Martini}} \bibnamefont{and}
  \bibinfo{author}{\bibfnamefont{E.}~\bibnamefont{C.~Monroe}},
  \emph{\bibinfo{title}{Experimental quantum computation and information}}
  (\bibinfo{publisher}{IOS Press}, \bibinfo{address}{Washington D.C.},
  \bibinfo{year}{2002}).

\bibitem[{\citenamefont{Basu et~al.}(2003)\citenamefont{Basu, Pollack, and
  Roy}}]{Basubook}
\bibinfo{author}{\bibfnamefont{S.}~\bibnamefont{Basu}},
  \bibinfo{author}{\bibfnamefont{R.}~\bibnamefont{Pollack}}, \bibnamefont{and}
  \bibinfo{author}{\bibfnamefont{M.~F.} \bibnamefont{Roy}},
  \emph{\bibinfo{title}{Algorithms in Real Algebraic Geometry}}
  (\bibinfo{publisher}{Springer-Verlag}, \bibinfo{address}{Berlin},
  \bibinfo{year}{2003}).

\bibitem[{mbc()}]{mbcr122007}
\bibinfo{note}{M. Bhattacharya and C. Raman, Phys. Rev. A \textbf{75}, 033405
  (2007); M. Bhattacharya and C. Raman, Phys. Rev. A \textbf{75}, 033406
  (2007).}

\bibitem[{sha()}]{shapiro2003}
\bibinfo{note}{E. A. Shapiro, I. Khavkine, M. Spanner and M. Yu. Ivanov, Phys.
  Rev. A \textbf{67}, 013406 (2003).}

\bibitem[{jes()}]{jessen2001}
\bibinfo{note}{G. Klose, G. Smith and P. S. Jessen, Phys. Rev. Lett.
  \textbf{86}, 4721 (2001).}

\end{thebibliography}
\end{document}